\documentclass{article}

\usepackage[nonatbib,final]{nips_2017Arxiv} 
\usepackage{graphicx}
\usepackage{authblk}

\usepackage[utf8]{inputenc} 
\usepackage[T1]{fontenc}    
\usepackage{hyperref}       
\usepackage{url}            
\usepackage{booktabs}       
\usepackage{amsfonts}       
\usepackage{nicefrac}       
\usepackage{microtype}      
\usepackage{soul}

\title{Using deep learning to reveal the neural code for images in primary visual cortex}

\begin{document}

\author[1,*]{ William~F.~Kindel}
\author[1]{Elijah~D.~Christensen}
\author[1,2]{Joel~Zylberberg}
\affil[1]{Department of Physiology and Biophysics, University of Colorado School of Medicine}
\affil[2]{Learning in Machines and Brains Program, Canadian Institute for Advanced Research}
\affil[ ]{*\textit{william.kindel@ucdenver.edu}}

\maketitle

\begin{abstract}
Primary visual cortex (V1) is the first stage of cortical image processing, and a major effort in systems neuroscience is devoted to understanding how it encodes information about visual stimuli. Within V1, many neurons respond selectively to edges of a given preferred orientation: these are known as \emph{simple} or \emph{complex} cells, and they are well-studied. Other neurons respond to localized center-surround image features. Still others respond selectively to certain image stimuli, but the specific features that excite them are unknown. Moreover, even for the simple and complex cells-- the best-understood V1 neurons-- it is challenging to predict how they will respond to natural image stimuli. Thus, there are important gaps in our understanding of how V1 encodes images. To fill this gap, we train deep convolutional neural networks to predict the firing rates of V1 neurons in response to natural image stimuli, and find that 15\% of these neurons are within 10\% of their theoretical limit of predictability. For these well predicted neurons, we invert the predictor network to identify the image features (receptive fields) that cause the V1 neurons to spike. In addition to those with previously-characterized receptive fields (Gabor wavelet and center-surround), we identify neurons that respond predictably to higher-level textural image features that are not localized to any particular region of the image.
\end{abstract}

\section{Introduction}

Our ability to see arises because of the activity evoked in our brains as we view the world around us. Ever since Hubel and Wiesel mapped the flow of visual information from the retina to thalamus and then cortex, understanding how these different regions encode and process visual information has been a major focus of visual systems neuroscience. In the first cortical layer of visual processing-- primary visual cortex (V1)-- Hubel and Wiesel identified neurons that respond to oriented edges within image stimuli. These are called \emph{simple} or \emph{complex} cells, depending on how sensitive the neurons' responses are to shifts in the position of the edge~\cite{hubel1959}. The simple and complex cells are well studied~\cite{lehky1992, david2004, montijn2016}. However, many V1 neurons are neither simple nor complex cells, and the classical models of the simple and complex cells often fail to predict how those neurons will respond to naturalistic stimuli~\cite{olshausen2005}. Thus, much of how V1 encodes visual information remains unknown. We use \emph{deep learning} to address this longstanding problem.

Recent advances in neural recording technology and machine learning have put  solving the V1 neural code within reach. Experimental technology for simultaneously recording from large populations of neurons-- such as multielectrode arrays-- has opened the door to studying how the collective behavior of neurons encodes sensory information. Moreover, methods of machine learning, inspired by the anatomy of the mammalian visual system, known as convolutional neural networks, have achieved impressive success in increasingly difficult image classification tasks~\cite{krizhevsky2012, lecun2015}. Recently, these artificial neural networks have been used to study the visual system~\cite{Yamins2016}, setting the state of the art for predicting stimulus-evoked neural activity in the retina~\cite{mcintosh2016} and inferior temporal cortex (IT)~\cite{yamins2014}. Despite these successes, we have not yet achieved a full understanding of how V1 represents natural images.

In this work, we present a convolutional neural network that predicts V1 activity patterns evoked by natural image stimuli. We use this network to predict the activity of 355 individual neurons in macaque monkey V1, and in doing so, this network represents the neural visual code for many neurons regardless of cell type. On held out validation data, the network predicts firing rates that are highly correlated ($\overline{r}_{\rm{model}} = 0.56 \pm 0.02$) with the neurons' actual firing rates. For 15\% of these neurons, the firing rates are predicted to within 10\% of the theoretical limit set by the trial-to-trial variation in the neural responses. To advance our understanding of the visual processing that takes place in V1, we invert the network to identify visual features that cause individual cells to spike. In the process, we identify novel functional cell types in monkey V1.

\section{Methods}
\label{meth}

\subsection{Experimental data}

\begin{figure}
\includegraphics[width=5.5in]{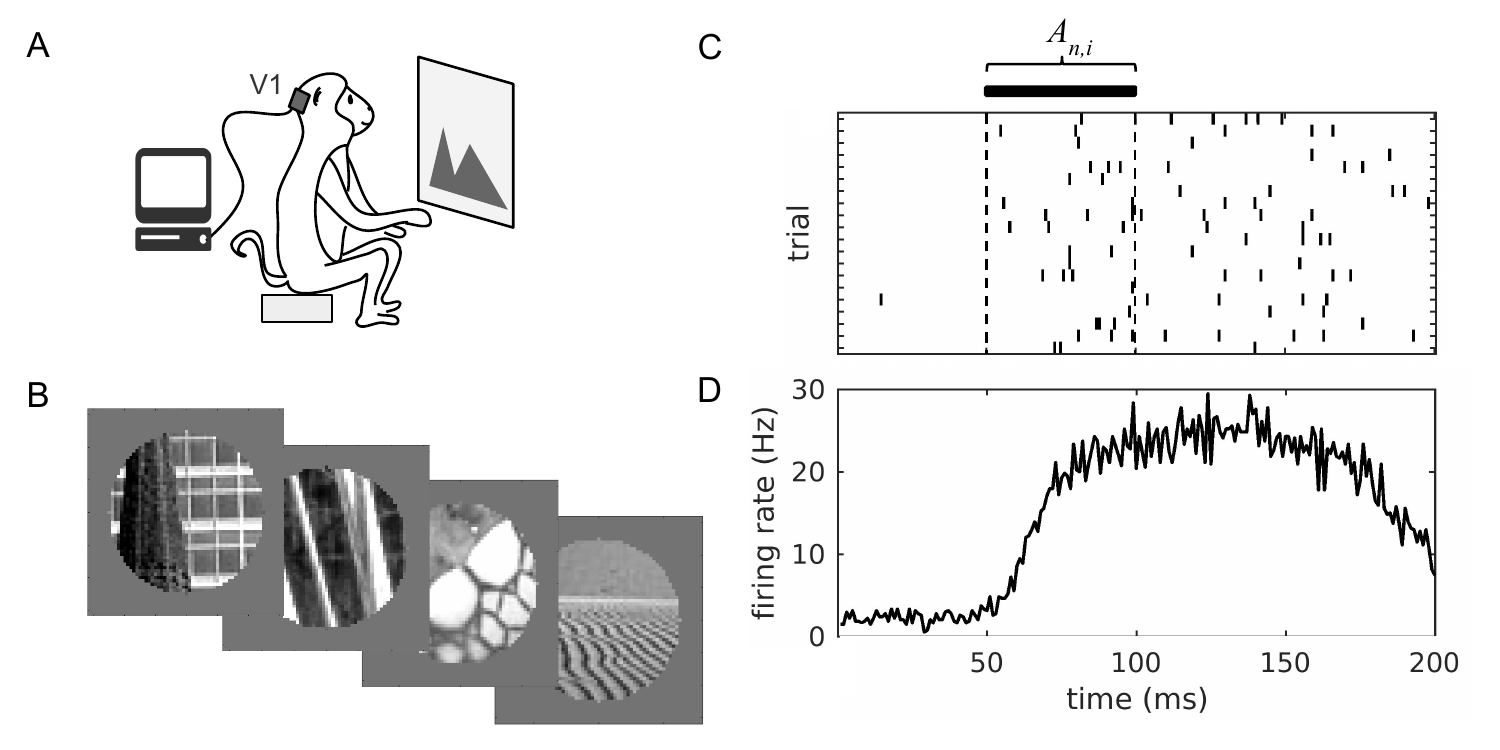}
\caption{Experimental data collection and processing. A) Neural activity was recorded in monkeys’ V1 as they were shown a series of images.  B) The image set contains 270 circularly cropped natural images.  C) The response of a single neuron over repeated presentations of an image. Ticks indicate the neuron's spiking; each row corresponds to a different image presentation trial. During the response window, the firing rate is computed and then averaged over trials to yield the average response $A_{n,i}$ used in our analysis. D)  The neuron responds to image stimuli with a $\sim 50$~ms latency from the image onset at $t = 0$, as seen in the peri-stimulus time histogram (firing rate plotted against time, averaged over all 270 images).}
\label{fig:data}
\end{figure}

We use publicly-available multielectrode recordings from macaque V1~\cite{coen2015}. In these experiments, macaque monkeys were anesthetized and then presented with a series of images, while the experimenters simultaneously recorded the spiking activity of a population of neurons in V1 (Fig.~1A,B) with a multielectrode array. These recordings were conducted in 10 experimental sessions with 3 different animals, resulting in recordings from a total of $392$ well-isolated neurons.   A full description of the data and experimental methods can be found in Coen-Cagli~et~al.~\cite{coen2015}. We use $37$ of these neurons from one session to determine how we construct our network (its hyper-parameters),  and the remaining $355$ neurons to evaluate its performance.  For each neuron $n$, we calculate the mean firing rate $A_{n,i}$ evoked by each image $i$, by averaging its firing rate across the 20 repeated presentations of that image. The firing rates are calculated over a window from 50 to 100~ms after the image was presented, to account for the signal propagation delay from retina to V1.

We analyze the responses to 270 natural images that are circularly cropped to the size of the retinal visual field (Fig.~1B). The full dataset contains responses to  natural and artificial stimuli  both full-sized and cropped. We use only natural images because we are interested in the real-world  behavior of the visual system, and we use only the cropped images because these images have the same visual field as the grating stimuli that we use to characterize the neurons as either orientation selective or not. Prior to training the neural network, we downsample the images using a non-overlapping 2~$\times$~2 window and crop them to a size of 33~$\times$~33 pixels.

\subsection{Deep neural network model}

\begin{figure}
\includegraphics[width=5.5in]{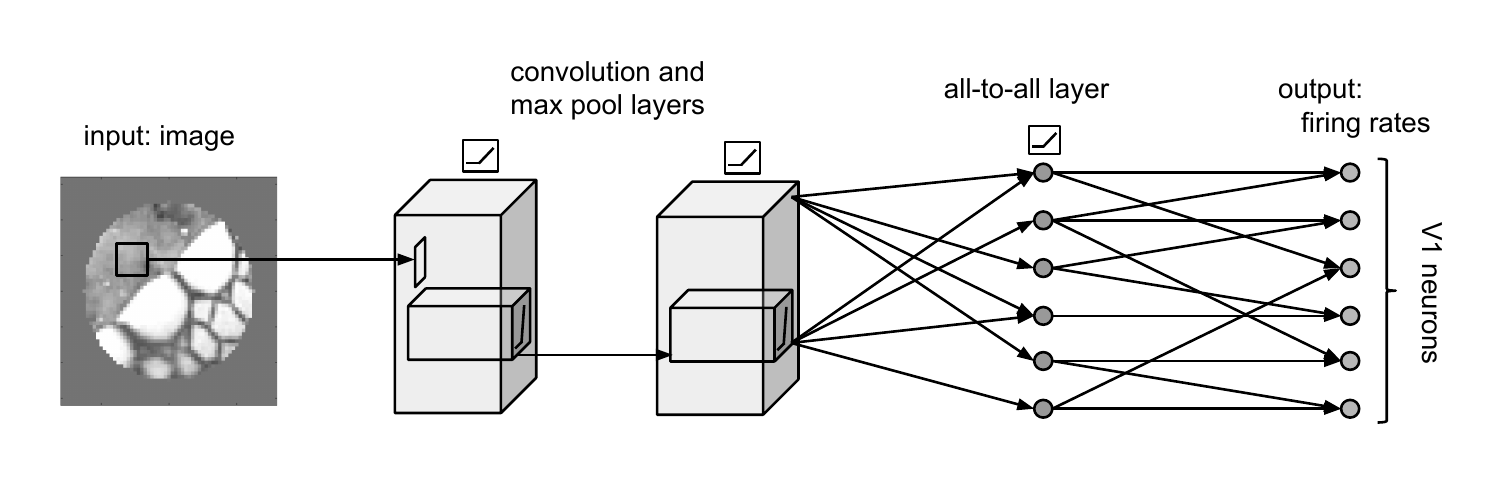}
\caption{The optimized architecture of the deep convolutional neural network model.  The network’s inputs are the pixel values of an image, and each output unit gives the predicted firing rate of a single neuron in monkey V1.}
\label{fig:model}
\end{figure}

To construct our predictive network, we use a convolutional neural network whose input is an image and whose output is the predicted firing rates of every neuron in a given recording session. As shown in Fig.~2, the network consists of a series of linear-nonlinear layers. The first layer(s) performs local feature extraction on the image by sweeping banks of convolutional filters over the image, and then applying a maximum pooling operation. These local features are then globally combined at the all-to-all layer(s) to generate the predicted firing rate for every neuron in that data session.

The number of each type of layer (convolutional with maximum pooling, or all-to-all), and the details about each layer (number of units, convolution stride, etc.), are optimized to maximize the accuracy of the neural activity predictions on the 37 neurons recorded in the second data session. We do this using a combination of manual and automated searches, where the results of our manual search inform the range of the hyper-parameter space for an automated random search~\cite{Bergstra2012}. Using the optimal parameters (Table~1) we train and evaluate our network with the remaining 9 datasets.



We train our network using a cross-validation procedure where we randomly subdivide a given dataset into a training subset (80\% of the images and corresponding V1 activity patterns) and an evaluation subset (20\% of the images). We then train all layers of our network using the TensorFlow Python package with the gradient-descent optimizer. We attribute a loss
\begin{equation}
L_n =  \frac{ \sum_i (y_{n,i} - A_{n,i} )^2 }{\rm{var}\it_i (A_{n,i})}
\end{equation}
to each neuron (indexed by $n$), where $i$ is the image index, $A_{n,i}$ is the measured response, and $y$ is the network's predicted response. The neurons' losses are summed yielding the total loss used by the optimizer. To ensure the performance generalizes, the training data is subdivided into data used by the optimizer to train the weights, and another small subset (14\% of the images) to stop the training when accuracy stops improving (early stopping).

To quantify the performance of the predictor, we compare the network’s predicted firing rates to the neurons' measured firing rates using a held-out evaluation set. This set is neither used to determine the hyper-parameters, nor to train the weights in our neural network. We calculate the Pearson correlation coefficient $r_{\rm model}$ between the predicted firing rates and the measured rates, for each neuron. To enable comparison with other work, we also calculate how much of the variance in the neurons' mean firing rates (over all test images) could be explained by the network's predictions: the fraction of variance explained (FVE).

\begin{table}[t]
\caption{Hyper-parameters of the optimized network.}
\label{sample-table}
\centering
\begin{tabular}{|ll|ll|}
 \toprule
  Conv. layer(s) & 2  & Dropout keep rate & 0.55 \\
  All-to-all layer(s) & 1 & &  \\
 \bottomrule  
\end{tabular}\\
\begin{tabular}{|ll|ll|ll|}
 \toprule
 \multicolumn{2}{|c|}{Conv. 1} & \multicolumn{2}{c}{Conv. 2} & \multicolumn{2}{|c|}{All-to-all}                  \\
 \midrule
 Num. filters & 16  & Num. filters & 32 & Num. elem. & 300     \\
 Conv. kernel     & 7  & Conv. kernel & 7  &  &   \\
 Maxpool stride     & 2 &     Maxpool stride     & 2 &   &     \\
 Maxpool kernel     & 3 &     Maxpool kernel     & 3  & &  \\
 \bottomrule
\end{tabular}
\end{table}

\subsection{Benchmarking the performance measures}

Because of the trial to trial variability in neural activity, no predictor could achieve $r_{\rm model} =1$. To understand how well our network can predict the V1 neurons' firing rates, we compare its predictability to a theoretical maximum value set by the variability in the neural responses. To compute this maximum, we generate fake data by drawing random numbers from Gaussian distributions with the same statistics as the measured neural data. For each neuron and image, we average over 20 of these such values to obtain a simulated prediction.  We then compute the correlation between these simulated predictions and the neurons' actual mean firing rates to find the maximum correlation $r_{\rm{max}}$ possible given the variability.


\subsection{Characterizing the selectivity of cells}

To interpret the breadth of our results, we group the cells into functional classes, and look at how well the firing rates from each class can be predicted by our neural network model. We classify cells by how selective they are to specific natural images, and by their selectivity to specific orientations of grating stimuli.

The selectivity of each neuron to specific natural images is quantified by the
\begin{equation}
\rm{image~selectivity~index} = \left( N - \frac{(\sum_i A_i)^2}{\sum_i(A_i^2)} \right)  \frac{1}{N - 1},
\end{equation}
where $A_i$ is the cell's firing rate indexed $i$ over the set of $N$ images~\cite{zylberberg2013}. This index has a value of zero for neurons that fire equally to all images, and a value of 1 for cells that spike in response to only one of the images.

The orientation selectivity is measured by the cell's
\begin{equation}
\rm{circular~variance} = 1 - \frac{\vert \sum_\theta A_\theta e^{i2\theta} \vert}{\sum_\theta  A_\theta},
\end{equation}
where $A_\theta$ is the neuron's firing rate in response to a grating oriented at angle $\theta$. The circular variance is less sensitive to noise than the more commonly-used orientation selectivity index~\cite{mazurek2014}. Following the results of Mazurek~et~al.~\cite{mazurek2014}, we use the thresholds: $\rm{circular~variance} < 0.6$ to define orientation selective cells (the simple and complex cells according to  the Hubel and Wiesel convention), and  $\rm{circular~variance} > 0.75$ for non-orientation-selective cells. We omit all other cells from these two groupings.

\subsection{Identifying visual features that cause the neurons to spike}

To use our model as a tool to investigate the functions of individual neurons, we use DeepDream-like~\cite{Mahendran2015} techniques to identify the visual features that cause each cell to spike. We invert our network by finding input images that cause a given cell to spike at a pre-specified level. To do this, we first take the fully trained network, and set Gaussian white noise images as the input. We then use backpropagation to modify the pixel values of the input image to push the chosen neuron's predicted firing rate towards the pre-specified level. Thus, we find an input image that induces the pre-specified response. We applied this procedure to several different neurons, and at several different target firing rates.

\section{Results}
\label{headings}

With our optimal network (Table~1), the predicted firing rates are highly correlated with the measured neuron's firing rates for most neurons (Fig. 3A) when evaluated on held-out data. The correlation between the predictions and actual neural firing rates is $\overline{r}_{\rm{model}} = 0.56 \pm 0.02$ ($\overline{\rm{FVE}}_{\rm{model}} = 0.36 \pm 0.01$) averaged over all 355 neurons in the evaluation set (Fig.~3B). Given the noise in the neural responses, a perfect predictor would achieve a correlation $r_{\rm{max}} = 0.851$. Therefore, our predictor achieves 66\% of the maximum possible performance.  Moreover, for many neurons, the predictability approaches this theoretical maximum: for 54 of the 355 cells, the predictability was within 10\% of the theoretical maximum. To show that our predictor performs complex, nonlinear processing on the images, we compare our results to a  linear model that predicts firing rates based on each pixel value of the input images. The linear model’s predictions have a correlation of $\overline{r}_{\rm{linear}} = 0.008 \pm 0.003$ to the measured firing rates, which is substantially lower than the network model (Fig.~3B).

\begin{figure}
\centering
\includegraphics[width=5in]{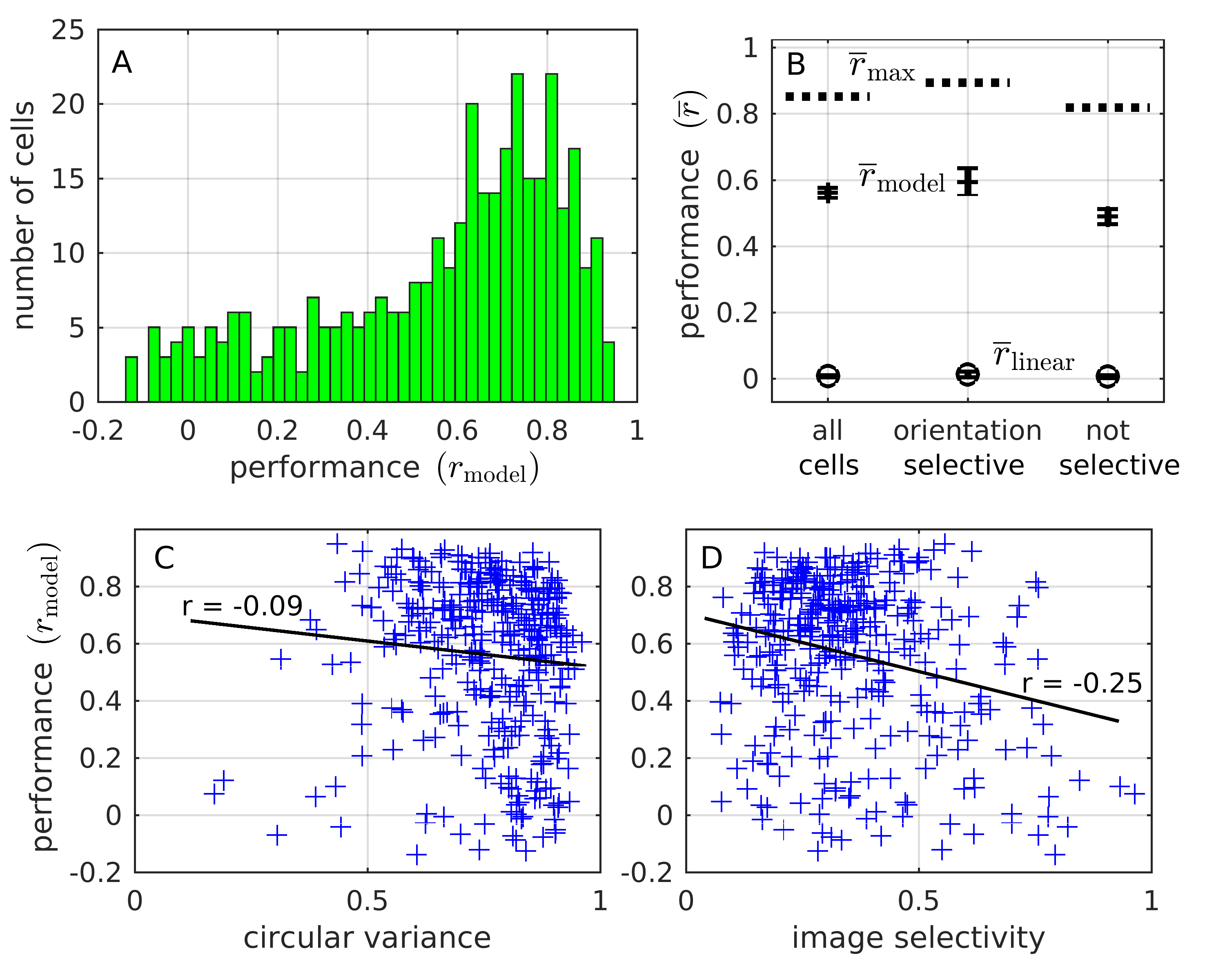}
\caption{The performance of the network model. A) A histogram of the Pearson correlation coefficient between the network predictions and the actual firing rates of all 355 neurons. These values are obtained using held-out data not used in training the network. B) The average performance of the predictor  $\overline{r}_{\rm{model}}$ (+) and the linear model $\overline{r}_{\rm{linear}}$ (o) for: 1) all neurons; 2) the orientation selective neurons; 3) non-orientation selective neurons. The limits on predictability $\overline{r}_{\rm{max}}$ are shown in dashed lines.  C) Scatter plot of how well the predictor can predict each neuron's firing rates (vertical axis) against the neuron's circular variance (horizontal axis). D) The predictability plotted against image selectivity index. For C and D, each data point corresponds to a single neuron.}
\label{fig:results}
\end{figure}

Because simple and complex cells have been extensively studied, we are motivated to compare the predictability of simple and complex cells to the predictability of the other neurons in the dataset.  Grouping the cells into orientation selective-- simple and complex-like cells-- and non-orientation-selective cells (see Methods), we find that our network predicts non-orientation-selective cell responses with $\overline{r}_{\rm{model}} = 0.49 \pm 0.02$, and orientation selective cell responses with $\overline{r}_{\rm{model}} = 0.60 \pm 0.04$ (Fig.~3B). Therefore, our model predicts the firing rates of both cell types, and performs slightly better on the simple and complex-like cells than the non-orientation-selective cells.

Given that some neurons' firing rates are well predicted by the network while others are not, we are motivated to ask what distinguishes predictable from unpredictable cells. To answer this question, we quantify the cells' orientation selectivity, and their image selectivity (see Methods).  Comparing the predictability of each cell's firing rates with its respective image selectivity index (Fig.~3D), and circular variance (Fig.~3C), we find that the predictability depends only weakly on these characteristics of the cells. Regardless of these values, some neurons' firing rates are well predicted while others are not. Thus, the orientation selectivity and image selectivity are only minor factors in determining how well our model performs.

\subsection{Identifying visual features that cause the neurons to spike}

By construction, our convolutional neural network assumes nothing about the neurons’ receptive fields. They are learned exclusively from the training data.  Therefore, we can use our network to determine the response properties of both well-characterized neurons (simple and complex cells) and poorly understood neurons in an unbiased manner. To do this, we invert the network and identify visual features that evoke specified responses in several of the well predicted neurons. Based on the measured firing rate distribution, we repeat this procedure for different target firing rates, from low (at the 20th percentile of the neuron's firing rate distribution) to high (80th percentile). 

\begin{figure}
\includegraphics[width=5.5in]{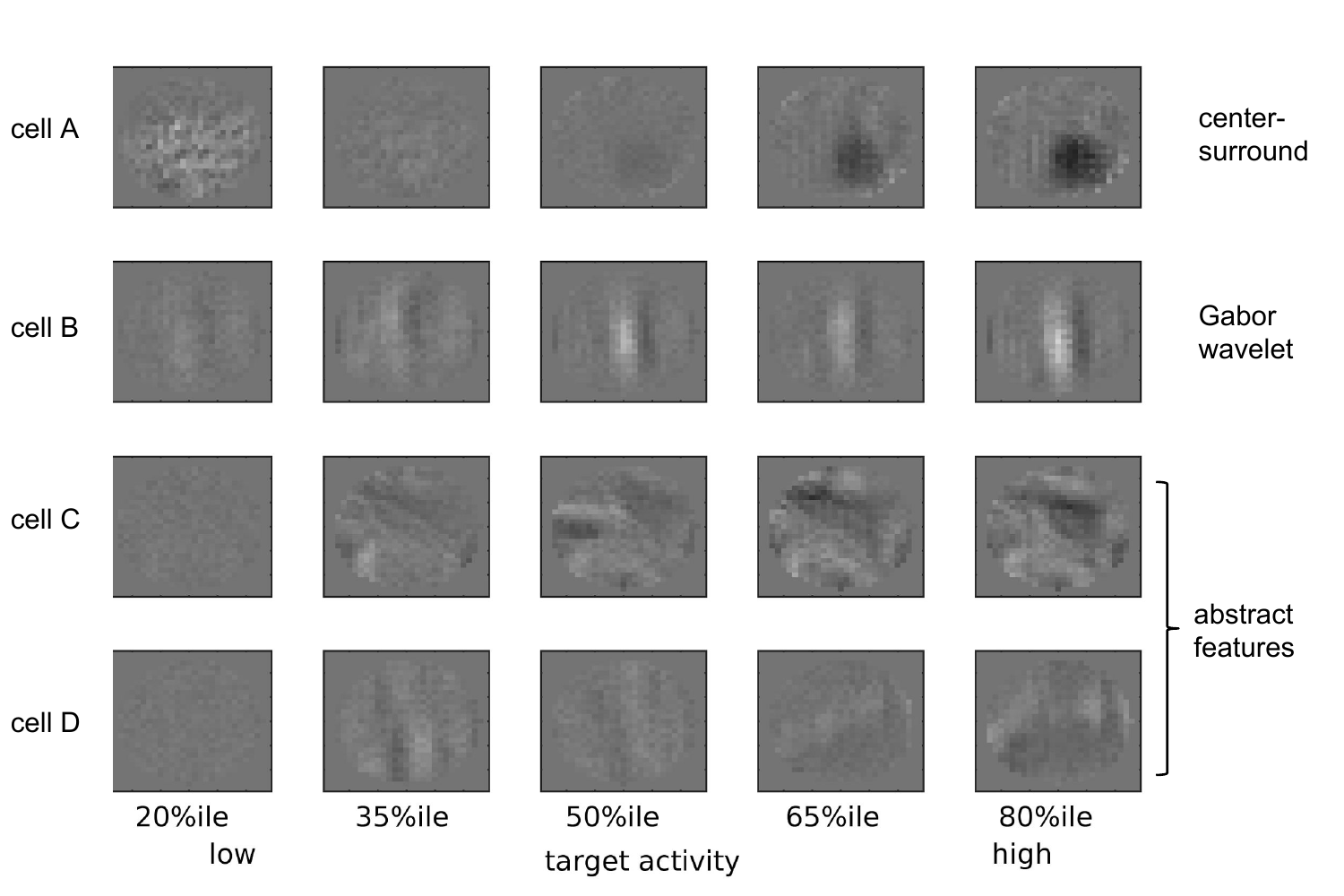}
\caption{Using the network model to reveal the visual features that drive individual neurons. For each neuron, we synthesize images that drove the predicted firing rates to the specified target values. These target firing rates are chosen to be different percentiles of the neurons' firing rate distribution. Cells A and B respond to localized image features, whereas cells C and D respond to more abstract image features.}
\label{fig:features}
\end{figure}

As seen in Fig.~4, this method allows us to classify several different types of cells, and identify novel response features. Cells A ($r_{ \rm model} = 0.88$) and B ($r_{\rm model} = 0.89 $) appear to function like previously characterized cells. Cell A responds to a center-surround image feature, and cell B's receptive field is a Gabor wavelet. In contrast, cells C ($r_{\rm model} = 0.91$) and D ($r_{\rm model} = 0.90$) respond to more abstract image features that are not well represented by simple localized image masks.

\section{Discussion}

We train a deep convolutional neural network to predict the firing rates of neurons in macaque V1 in response to natural image stimuli.  We find that the firing rates of both orientation-selective, and non-orientation-selective neurons can be predicted with high accuracy. Moreover, we find that the network can identify the image features that cause the neurons to spike. This procedure reveals both canonical localized neural receptive fields (such as Gabor wavelets and center-surrounds), as well as abstract image features (previously uncharacterized in V1 neural receptive fields) that were not localized to a single region of the image. Our results have implications for developing new computer vision algorithms as well as studying the visual centers of the brain. 

\subsection{Comparisons to other work}


Studying visual processing in V1 we find that the optimal architecture of our convolutional neural network is relatively shallow compared to recent results by Yamins~et~al.~\cite{yamins2014}. Deep neural networks require large training datasets to generalize~\cite{srivastava2014}. With only 270 images to train and evaluate our network, the optimal architecture that we find is likely a balance between the architecture that truly represents V1 encoding given infinite data and one that generalizes well. To probe whether a deeper network could more truly represent V1, a much larger dataset is required; this highlights the persistent limitation that small datasets impose on deciphering the neural code.

Although, it is difficult to fairly compare the performances of published results for a variety of factors, we predict neural activity with performance that is comparable to the state of the art. Over all neurons, the correlation between our network predictions and the actual neural firing rates is $\overline{r}_{\rm{model}} = 0.56 \pm 0.02$. For comparison, Lau~et~al.~\cite{lau2002} achieved predictability of $r = 0.45$ for simple cells and $r =0.31$ for complex cells, and Prenger~et~al.~\cite{prenger2004} achieved $r = 0.24$ averaged over all cells. Lehky~et~al.~\cite{lehky1992} achieved $r = 0.78$ and Willmore et al.~\cite{Willmore2010} achieved a predictability of $\rm FVE = 0.4$. However, some contextual factors confound direct comparison to these results. Specifically, Lehky~et~al. selected neurons that are easier to predict by specifically choosing neurons that responded strongly to the presentation of bars of light, and Willmore~et~al. adjusted their image to match the respective field of each neuron they predicted. Despite methodology differences, we report comparable performance or better to recent published results.

\subsection{Implications for machine learning}


While supervised learning methods lead to impressive performance on image categorization tasks~\cite{krizhevsky2012, lecun2015}, the trained networks are easily fooled by imperceptible image manipulations~\cite{nguyen2015}, and they require large amounts of training data to achieve high levels of performance. Primates, however, are not so easily fooled. They can learn to perform classification tasks given only small numbers of training examples. Thus, it may be possible to improve the deep networks used for computer vision by building on the primate brain's representation. Deep neural networks -- like the one presented here -- could be pre-trained to predict primate V1 firing patterns, and subsequently trained to perform object recognition tasks. Thus, combining our approach with traditional learning could lead to more robust and data-efficient algorithms.

\subsection{Implications for neuroscience and medicine}

By inverting our network (Fig.~4), we show that we can use the network as a tool to investigate the neurons' response properties. As demonstrated by distinguishing Gabor wavelet (cell B) from center-surround (cell A) receptive fields, this tool can identify and classify functional cell types. Going forward, this tool shows promise for characterizing the response properties of more cells in V1, and precisely defining functional cell types that were previously overlooked.  Looking beyond V1, these methods could be applied to understanding higher level cortical processing, such as visual encoding in V2. By finding the features that elicit a response in V2 neurons, this tool could help fill the visual encoding knowledge gap~\cite{Ziemba2016} that exists between the abstract encoding of IT and V4 and the low-level encoding of the retina and V1.

Additionally, our results have taken a key step towards cracking the neural code for how visual stimuli are translated into neural activity in V1. This would be a major step forward in sensory neuroscience, and would enable new technologies that could restore sight to the blind. For example, cameras could continuously feed images into networks that would determine the precise V1 activity patterns that correspond to those images: a camera to brain translator. Brain stimulation methods like optogenetics~\cite{ozbay2015} could then be used to generate those same activity patterns within the brain, thereby restoring sight. 

To master neural encoding, we propose closing the loop between conducting the experiments and performing analysis.  By using the network to generate visual stimuli hypothesized to evoke particular patterns of neural activity, experiments could directly probe the neural code, and in doing so pave the way for a new class of neuroscience experiments.
%
 

\subsubsection*{Acknowledgments}

We thank Adam Kohn and Ruben Coen-Cagli for providing the experimental data, and we thank Gidon Felson, John Thompson, Adam Kohn and Ruben Coen-Cagli for providing invaluable feedback. This research is supported by the National Institutes of Health under award numbers T15 LM009451 and T32 GM008497, the Canadian Institute for Advanced Research (CIFAR) Azrieli Global Scholar Award, and the Google Faculty Research Award.


\bibliography{V1_predictorbib17may}{}
\bibliographystyle{unsrt}
\end{document}